\documentclass{PoS}

\usepackage[english]{babel}
\usepackage[ansinew]{inputenc}
\usepackage{amsthm, amssymb, amsmath, latexsym}
\usepackage{graphicx, graphics, subfigure}
\usepackage{multirow}
\usepackage{slashed}

\usepackage{ragged2e} 
\usepackage{marvosym} 

\newcommand{\Dov}    {\mathcal{D}_\text{ov}}

\usepackage[square,comma,numbers]{natbib}
\usepackage{float}

\usepackage{cleveref}

\graphicspath{{pics/}}

\title{The phase structure of a chirally-invariant Higgs-Yukawa model }
\ShortTitle{The phase structure of a chirally-invariant Higgs-Yukawa model }

\author{Prasad Hegde$^{a}$, George~W.-S.~Hou$^{a}$,
Karl~Jansen$^{b}$, \speaker{Bastian~Knippschild}$^{\,\,d}$,\newline
C.-J. David~Lin$^{e,h}$, Kei-Ichi~Nagai$^{f}$, Attila~Nagy$^{b,c}$, Kenji~Ogawa$^{g}$\\\\
	$^{a}$ Department of Physics, National Taiwan University, Roosevelt Road, Taipei 10617, Taiwan\\
	$^{b}$ NIC, DESY, Platanenallee 6, Zeuthen D-15738, Germany\\
	$^{c}$ Institut f\"ur Physik, Humboldt-Universit\"at zu Berlin, D-12489 Berlin, Germany\\
	$^{d}$ HISKP, Rheinische Friedrich-Wilhelms-Universit\"at, D-53115 Bonn, Germany\\
	$^{e}$ Institute of Physics, National Chiao-Tung University, Hsinchu 300, Taiwan\\
	$^{f}$ Kobayashi-Maskawa Institute, Nagoya University, Nagoya, Aichi 464-8602, Japan\\
	$^{g}$ Department of Physics, Chung-Yuan Christian University, Chung-Li 32023, Taiwan\\
	$^{h}$ Division of Physics, National Centre for Theoretical Sciences, Hsinchu 300, Taiwan\\\\
	E-mail: \email{b.knippschild@gmx.de}\\\\}

\abstract{
\vspace{+0.2cm}
We present new results of our ongoing project on the investigation of the
phase structure of the Higgs-Yukawa model at small and large bare Yukawa couplings. The critical exponents of the second order bulk phase transitions of this model are determined from finite-size analyses and compared to the pure O(4)-model to test for triviality and the possibility of having a non-Gaussian fixed point. In addition, we will present a first study of Higgs boson masses and fermion correlation functions. 
}

\FullConference{31st International Symposium on Lattice Field Theory - LATTICE 2013\\July 29 - August 3, 2013\\Mainz, Germany}

\begin{document}

%
%
\section{Introduction}

The Standard Model (SM) was completed with the finding of the Higgs boson in the ATLAS and CMS experiments at CERN in 2012 \cite{Aad:2012tfa,Chatrchyan:2012ufa}. However, a lot of questions cannot be answered within the SM like the amount of CP-violation or dark matter and energy. We investigate the phase structure of a chirally-invariant Higgs-Yukawa (HY) model which could lead to a natural extension of the SM in case of a large renormalized Yukawa coupling. In this scenario bound states of fermions would be possible which are coupled through a Higgs boson exchange. For comparison and validation of our analysis methods we compare two bulk phase transitions in the HY model at small and large bare Yukawa couplings with the bulk phase transition found in the O(4) model. 

The action of the O(4) model is given by
\begin{equation}
S_B[\Phi] = -\kappa \sum\limits_{x,\mu} \Phi_x^{\dagger} \left[\Phi_{x+\mu} + \Phi_{x-\mu}\right] + \sum\limits_{x} \Phi_x^{\dagger} \Phi_x + \hat{\lambda}\sum\limits_{x} \left[ \Phi_x^{\dagger} \Phi_x - 1 \right]^2
\end{equation}
where the scalar field $\Phi$ is a real four-component vector, $\kappa$ the hopping parameter and $\hat{\lambda}$ the quartic self coupling. Connection to continuum formulation can be made via
\begin{equation}
\varphi = \sqrt{2 \kappa} \left( \begin{array}{c} \Phi^2 + i\Phi^1 \\ \Phi^0 - i \Phi^3 \end{array} \right) ,\quad \lambda_0 = \frac{\hat{\lambda}}{{4 \kappa^2}},\quad m_0^2 = \frac{1 - 2 \hat{\lambda} -8 \kappa}{\kappa}.
\end{equation}
On top of the O(4)-model the interaction of the Higgs-boson field with a degenerate heavy fermion doublet is described by 
\vspace{-0.25cm}
\begin{equation}
S_\Psi = \sum_{x,x'}\bar{\Psi}_x \left[\Dov + yP_+\Phi^\alpha\theta_\alpha^\dagger\ \hat{P}_+ + yP_-\Phi^\alpha\theta_\alpha\hat{P}_-\right]_{x,x'}\Psi_{x'}
\vspace{-0.25cm}
\end{equation}
with $\theta_{1,2,3} = -i\tau_{1,2,3}$, $\tau$ the Pauli matrices, $\theta_4=1_{2\times 2}$, the chiral projectors $P_\pm$ and $\hat P_\pm$, and $y$ the Yukawa coupling. The free-fermion part is described by the overlap operator $\Dov$ \cite{Neuberger:1997fp} which is usually numerically challenging. Gauge fields are not included in this model so $\Dov$ is analytically known and its numerical computation is feasible \cite{Luscher:1998pqa}. The Polynomial Hybrid Monte Carlo algorithm \cite{Frezzotti:1997ym,Gerhold:2010wy} was used to create configurations of Higgs boson fields.

The magnetisation of the Higgs-boson field
\vspace{-0.25cm}
\begin{equation}
 m_L = V^{-1}\left\langle \left|\frac{1}{V}\sum_{x} \Phi_x \right| \right\rangle  \vspace{-0.25cm}
\end{equation}
serves as an order parameter for the phase structure investigation. To ensure a non-vanishing magnetisation in the broken phase the $\Phi$ field is rotated which is equivalent of including an external field $J$ and taking the limit $J \rightarrow 0$ \cite{Hasenfratz:1989ux,Gockeler:1991ty}.

\begin{figure}[H]
\begin{center}
\includegraphics[width=5.3cm]{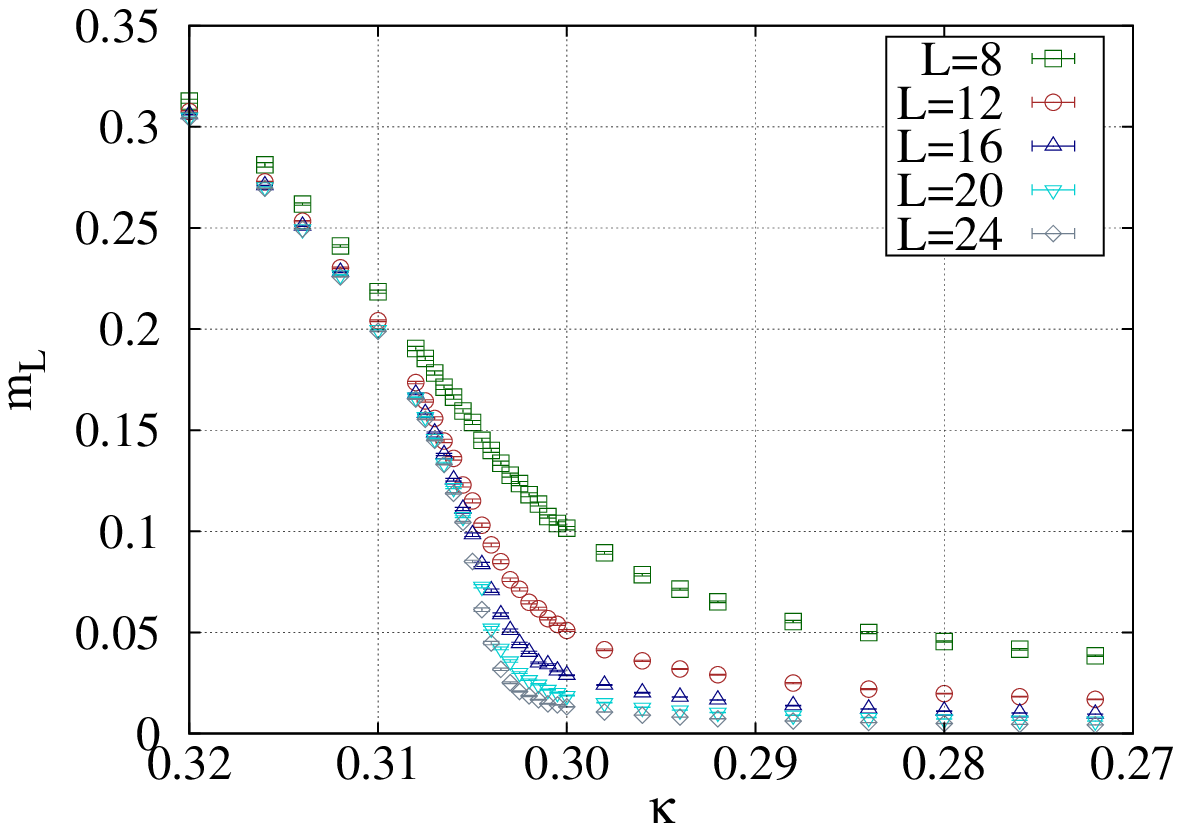}
\hspace*{-0.6cm}
\includegraphics[width=5.3cm]{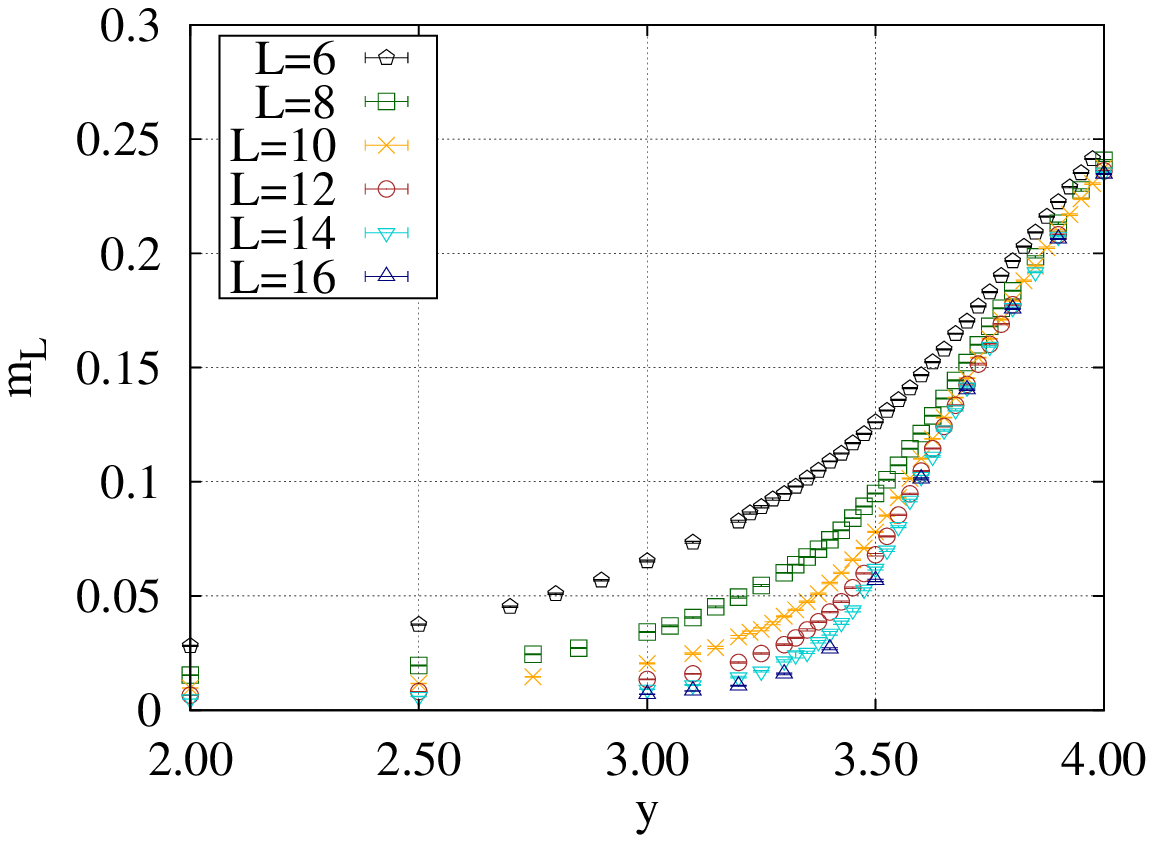}
\hspace*{-0.6cm}
\includegraphics[width=5.3cm]{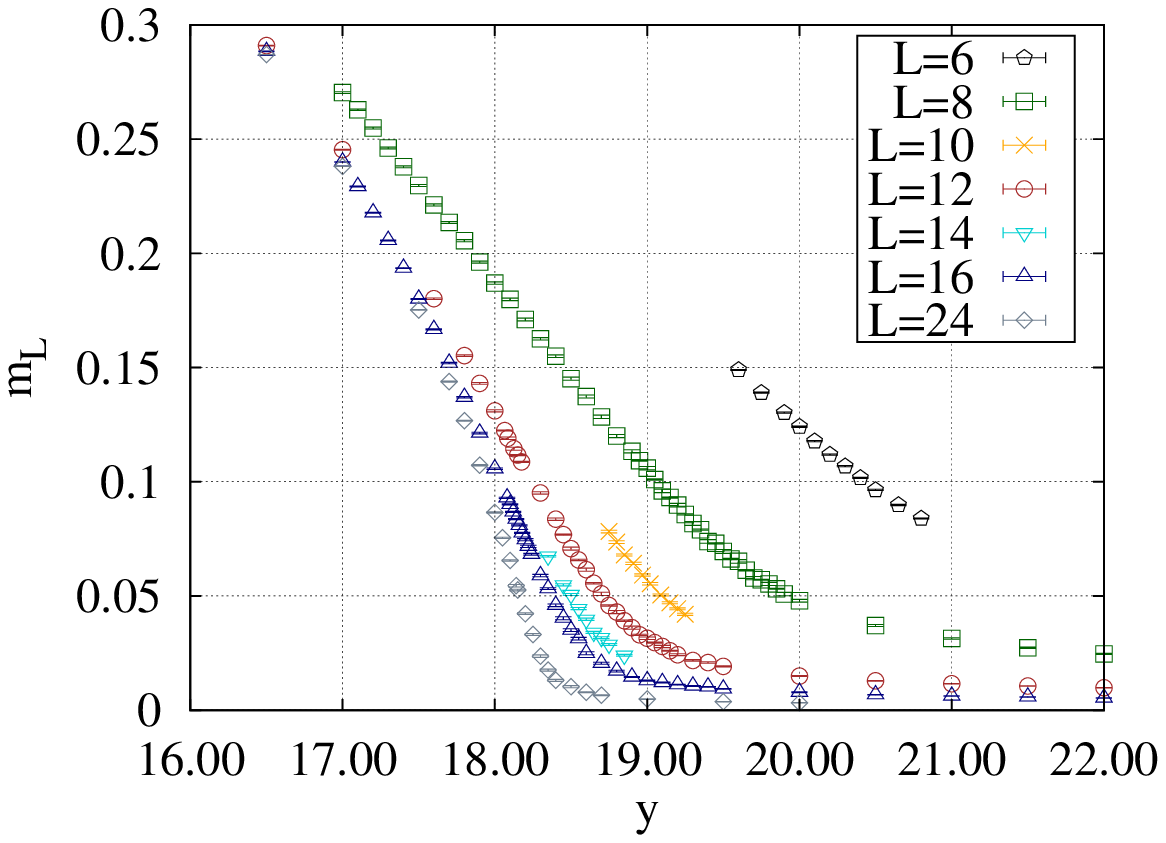}
\end{center}
\label{Magnetisations:FIG}
\vspace{-0.6cm}
\caption{Magnetizations in the O(4)-model {\it(left)}, and the HY model at small {\it(middle)} and large {\it(right)} Yukawa couplings for various volumes.}
\end{figure}

The magnetisation is shown in fig.\,\ref{Magnetisations:FIG} for the O(4)-model and the two phase transitions in the HY model at small and large Yukawa couplings respectively. While the phase transition in the O(4)-model is scanned in the hopping parameter $\kappa$ the phase transition of the HY model is scanned in the Yukawa coupling $y$ by simultaniously fixing $\kappa=0.06$. Other values of $\kappa$ were investigated as well \cite{Bulava:2011jp,Bulava:2012rb,Bulava:2012bc} but are not reported here. A scan in $\kappa$ in the HY model is currently under investigation. Bulk phase transitions can be observed in all three scenarios where the magnetisation is non zero in the broken phase and goes to zero in the symmetric phase. The magnetisation is not exactly zero in the symmetric phase due to finite volume effects. The absence of jumps in the magnetisation is evidence of a second order phase transition.
\vspace{-0.40cm}
\vspace{-0.8cm}

%
%
\section{Analysis methods}
Finite-size scaling (FSS) is used to investigate the continuum behaviour (universality class) of phase transitions. The idea of FSS is that finite volume scaling behaviour is connected to the critical exponents (anomalous dimension of operators) of the phase transition in infinite volume. The investigation of the magnetisation indicates such a second order phase transition. In each universality class the critical exponents are unique and hence the continuum limit of the HY model and the O(4)-model can be compared to each other.

In this analysis the focus lies on the investigation of FSS of the magnetic susceptibility
\vspace{-0.25cm}
\begin{equation}
\chi_L = V \left[ \langle m^2_L \rangle - \langle m_L\rangle^2\right].
\vspace{-0.25cm}
\end{equation}
FSS predicts different behaviour of the susceptibility in various bare parameter regions \cite{Barber1984}:
\vspace{-0.15cm}
\begin{align}
\chi_L\left(|T-T_c^L| \gg 1\right) &\sim \left|T-T_c^L\right|^{-\gamma},\label{FSS1}\\
\chi_L\left(|T-T_c^L| \rightarrow 0\right) &\sim L^{1/\nu},\label{FSS2}\\
T_c^L-T_c^\infty&\sim L^{-1/\nu},\label{FSS3}
\vspace{-0.5cm}
\end{align}
where $T$ is the scanning parameter, $\kappa$ or $y$. In \cref{FSS1} the behaviour of $\chi_L$ is described far away from the phase transition. Volume effects become negligible in this region and it is scanned in the Higgs-boson mass. Hence, susceptibility scales with the anomalous dimension of field renormalization, $\gamma$, which is not investigated here. \Cref{FSS2,FSS3} describe FSS very close to the bulk phase transition where volume effects become large. The only relevant operator in the continuum limit is the mass operator so susceptibility's height and position scales with the volume and the anomalous dimension of mass, $\nu$. 

The O(4)-model has a Gaussian fixed point \cite{Luscher:1988uq} and hence $\nu$ is known to be $1/2$. Due to triviality, the FSS proposed in \cref{FSS1,FSS2,FSS3} must be modified by logarithmic corrections \cite{Gockeler:1992zj}:
\vspace{-0.15cm}
\begin{align}
\chi_L\left(|T-T_c^L| \rightarrow 0\right) &\sim L^{2}\log(L)^{1/2},\\
T_c^L-T_c^\infty&\sim L^{-2}\log(L)^{-1/2}.
\vspace{-0.25cm}
\end{align}
Similar logarithmic corrections would also appear in the HY model in case of triviality but the corresponding exponent of the logarithm was never computed analytically to the best of our knowledge. It is not necessary that these corrections are identical to the ones in the O(4)-model because fermion loops will appear. This behaviour will be investigated in the analysis presented in the next paragraphs.

Various methods were used to extract the critical exponent $\nu$ from the scaling behaviour of susceptibility described in \cref{FSS1,FSS2,FSS3}. Our standard analysis is a global fit to all volumes simultaneously \cite{Jansen:1989gd} which respects the FSS:
\vspace{-0.25cm}
\begin{align}
\chi_L\left(T; \xi\right) &=\nonumber A_1\left( \left[ L^{2}(\log{L})^\xi\right]^{-1/\nu}+ A_{2,3}\cdot\tau^2 \right)^{-\gamma/2}\nonumber\\
\vspace*{-.2cm}
\tau &= \left(T-T_c^L\right) = \left[T-\left(T_c^\infty+C\cdot \left[L^{-1}\cdot(\log{L})^{-\xi/2}\right]^{b}\right)\right].\label{globalfit}
\vspace*{-0.25cm}
\end{align}
This fit function allows for a direct determination of the critical exponents $\nu$ and $\gamma$, and the phase transition point in infinite volume, $T_c^\infty$. It has eight free fit parameters: $A_1, A_2, A_3, C, \nu, \gamma, T_c^\infty$, where the exponent $b$, which describes the peak shift of susceptibility, is directly related to the anomalous dimension of the mass operator $b=1/\nu$ but it is left free for generality. 

The logarithmic corrections for FSS are also included in \cref{globalfit}. It is not possible leaving the logarithmic exponent $\xi$ as a free fit parameter because it always appears in combination with the critical exponent $\nu$. Hence, $\xi$ must be fixed in each individual fit. The strategy is to scan through various values of $\xi$ to find $\nu$ in agreement with $1/2$, its value in case of triviality.   

\begin{figure}[H]
\begin{center}
\includegraphics[width=5.3cm]{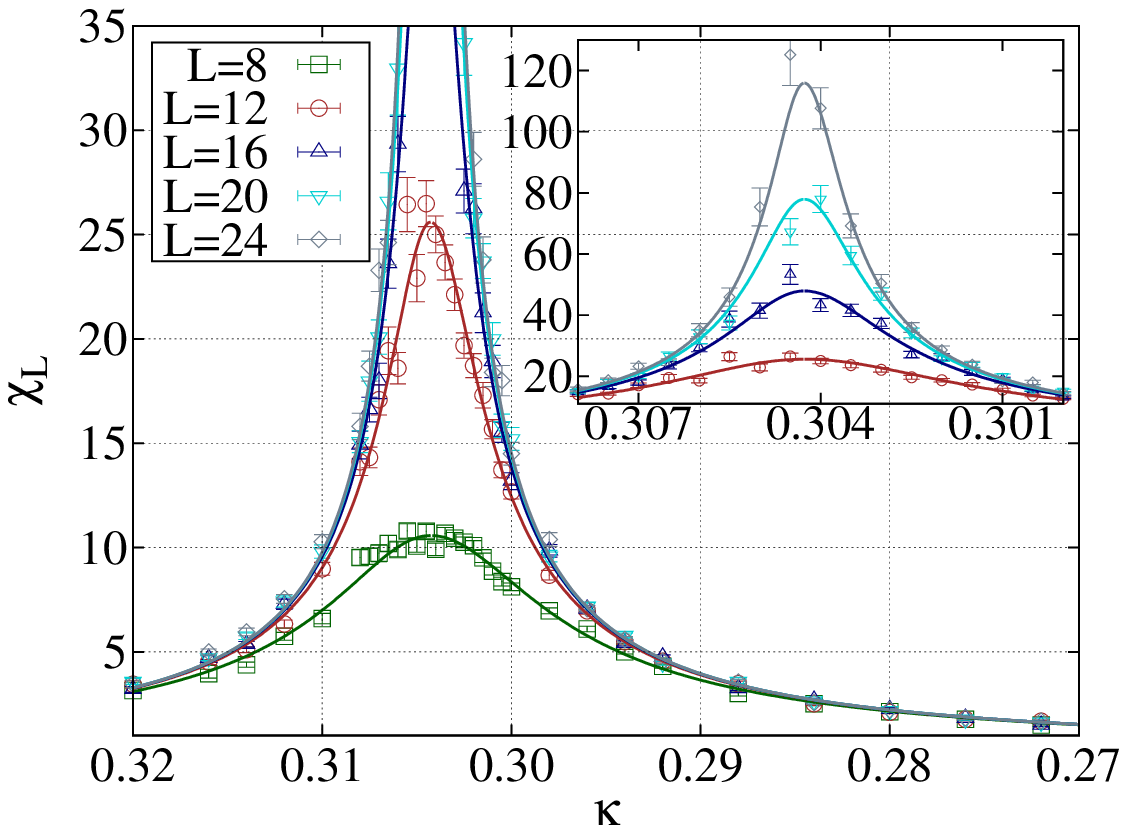}
\hspace*{-0.6cm}
\includegraphics[width=5.3cm]{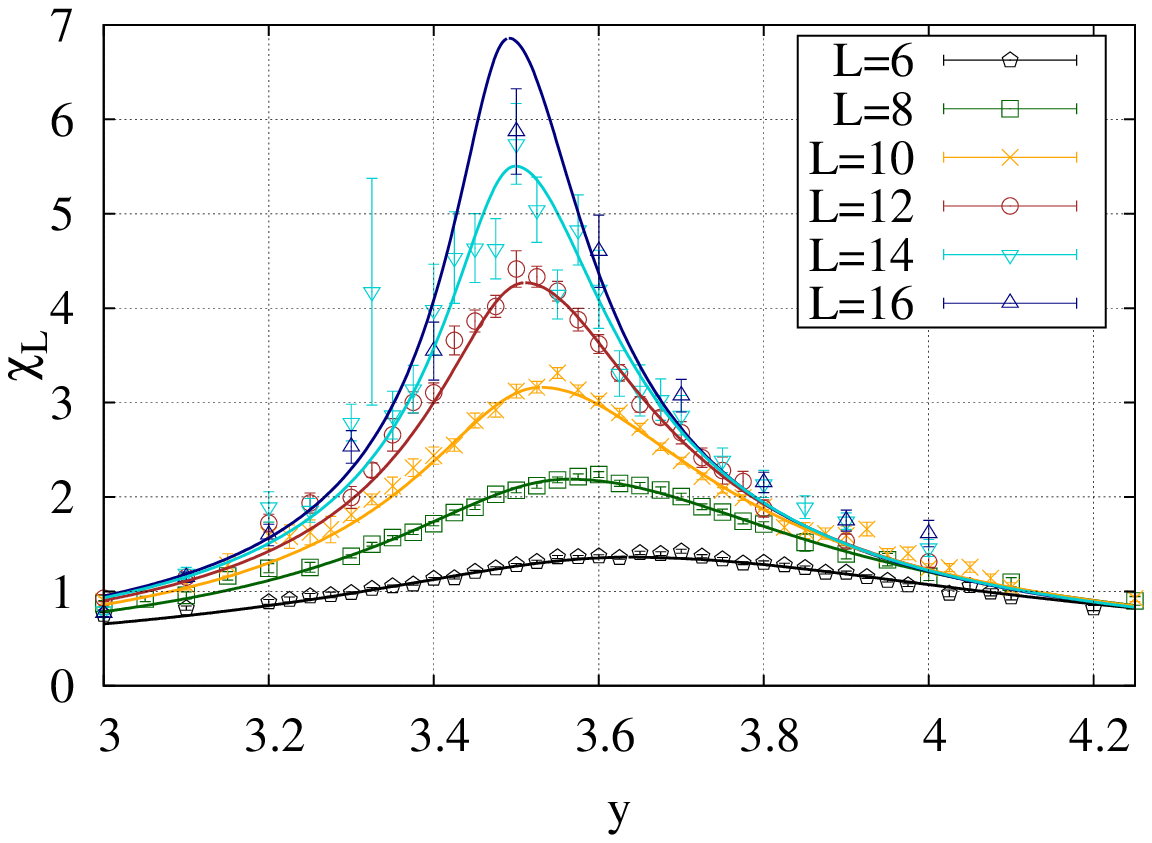}
\hspace*{-0.6cm}
\includegraphics[width=5.3cm]{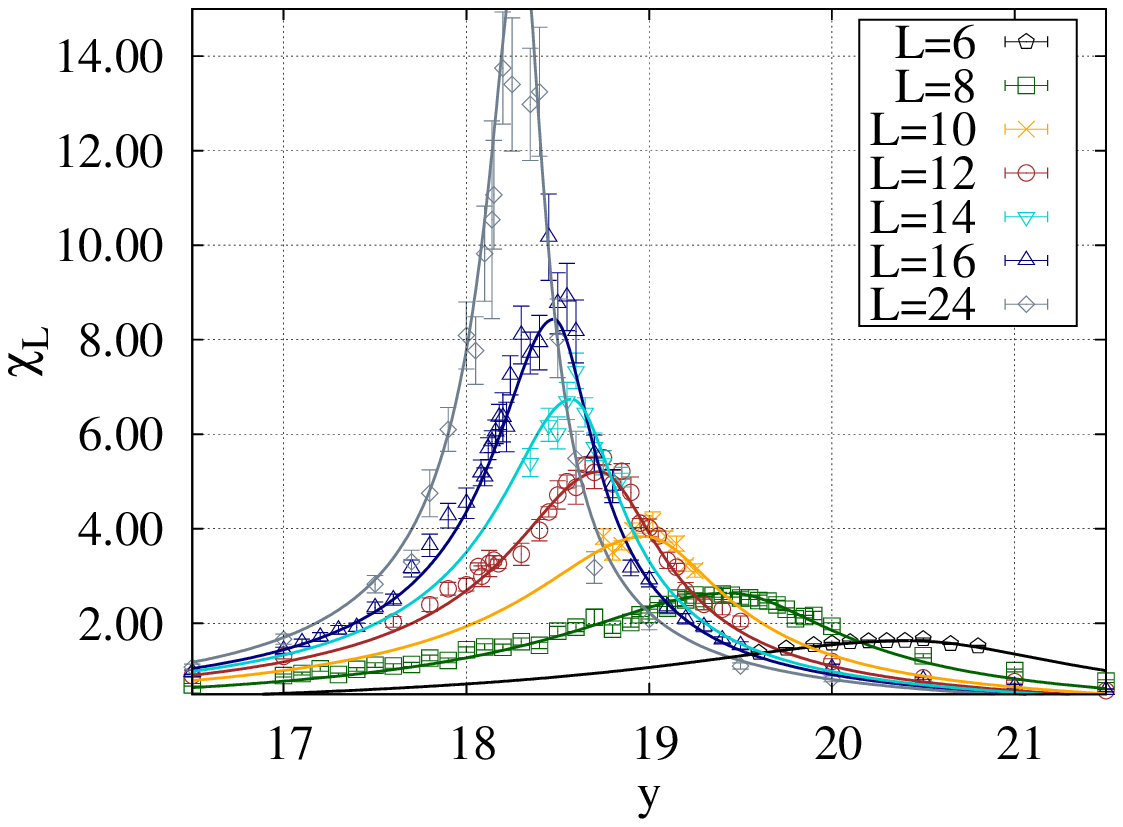}
\vspace{-0.85cm}
\caption{Susceptibilities in O(4)-model {\it (left)} and the HY model in the small {\it (middle)} and large {\it (right)} bare Yukawa coupling region. The curves are fits to \cref{globalfit} with $\xi=0$.}
\end{center}
\label{Susceptibility:FIG}
\end{figure}

The magnetic susceptibility and fit results from \cref{globalfit} with $\xi=0$ are shown in fig.\,\ref{Susceptibility:FIG} for the O(4) and HY model. The fit covers the data well which is very strong evidence for a second order phase transition. While the volume dependence of the peak position is very prominent in the HY model it is very small in the O(4) model. In fact, the shift is so small in the O(4) model that we cannot resolve it in our fits with the given statistics. 
\vspace{-0.65cm}

To address systematic effects, two additional fit methods were taken into account. In both methods the susceptibility is fitted for each volume individually. An additional step is necessary to extract $\nu$ either from the peak height \cref{FSS2} or the peak position \cref{FSS3}. The susceptibility in the O(4)-model does not show any shift of the peak position. Furthermore, a fit based in \cref{FSS3} would lead to a non-linear fit function. That is the reason that only the peak height is considered to extract $\nu$. \Cref{FSS2}, which describes the volume dependence of the peak height, leads to the linear fit function 
\begin{equation}
\chi_L^\text{max}(\xi)=A_1\cdot (L [\log{L}]^\xi)^{1/\nu}
\label{extractionofnu}
\end{equation}
with two free fit parameters $A_1$ and $\nu$. As for the global fit the logarithmic correction is included. The exponent $\xi$ is not a free fit parameter and must be kept fixed in each fit. With respect to the logarithmic correction the fit strategy is the same as for the global fits.

One of the fit functions to extract the susceptibility's height was proposed in \cite{Jansen:1985nh}. The fit function is given by 
\vspace{-0.15cm}
\begin{equation}
\chi_L(T) = a + c\cdot T + \frac{d}{1+e\cdot\left| T-T_c^L\right|^g},
\label{fitvolume}
\end{equation}
\vspace{-0.15cm}
with 6 free fit parameters: $a, c, d, e, g, T_c^L$. The other fit method is a naive fit of quadratic form
\begin{equation}
\chi_L(T) = m + p\cdot T + q\cdot T^2,
\label{fitquadratic}
\vspace{-0.25cm}
\end{equation}
with three free fit parameters $m, p, q$. It can only be performed very close to the phase transition.

\begin{figure}[H]
\begin{center}
\includegraphics[width=5.3cm]{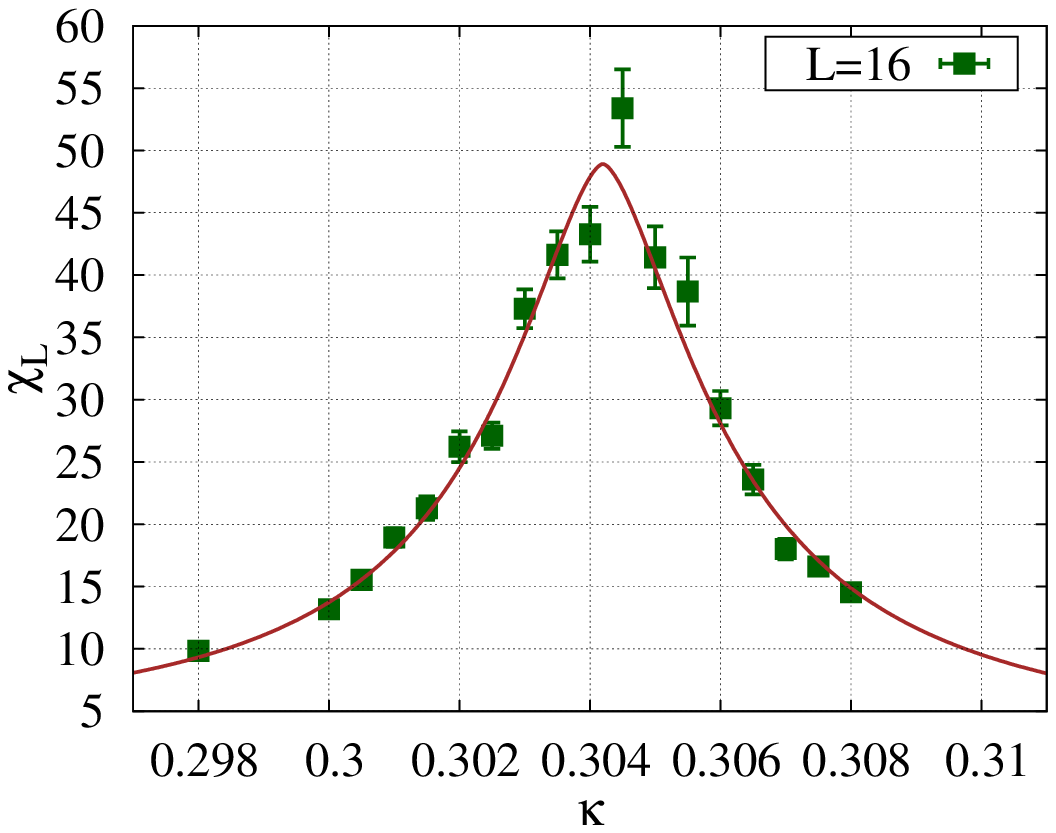}
\hspace*{-1.0cm}
\includegraphics[width=5.3cm]{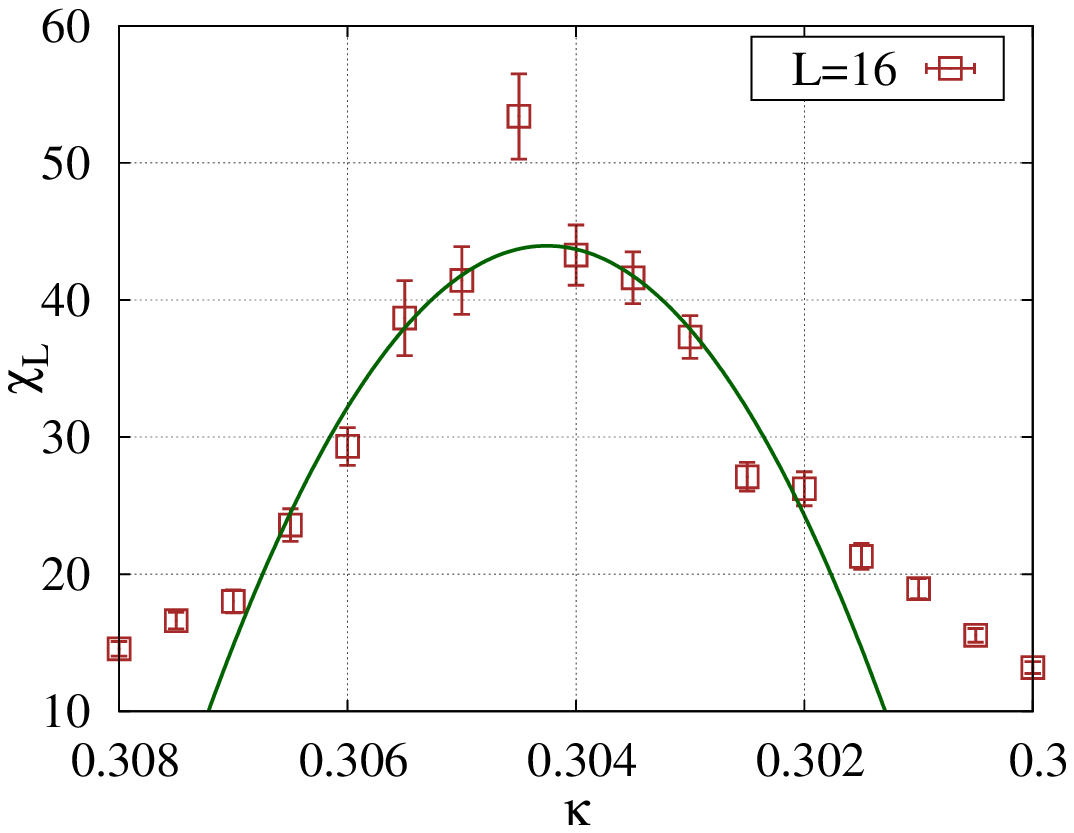}
\hspace*{-0.5cm}
\includegraphics[width=5.3cm]{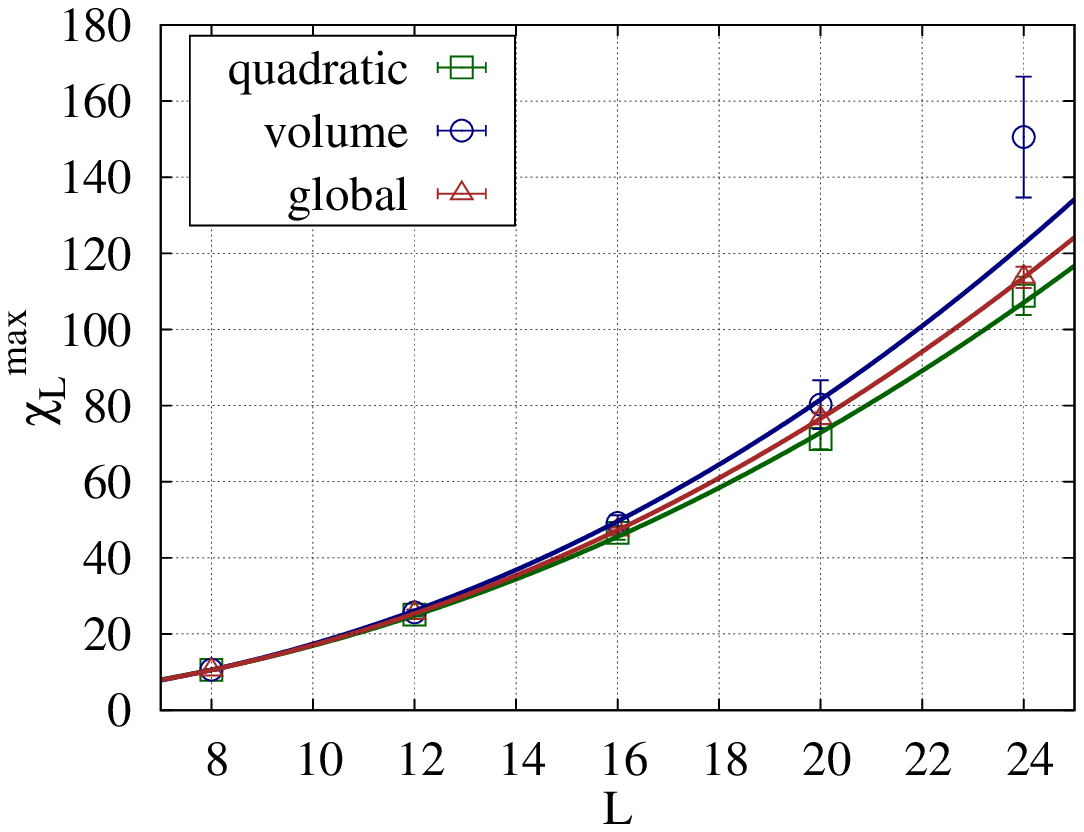}
\vspace{-0.40cm}
\caption{Individual volume fits to $L=16$ in the O(4)-model to \cref{fitvolume} {\it(left)} and \cref{fitquadratic} {\it(middle)}. Extraction of $\nu$ from \cref{extractionofnu} for the three fit methods {\it(right)}.}
\label{alternativefits:FIG}
\end{center}
\end{figure}

Exemplary fit results of the two methods are shown in the left and middel plot of fig.\,\ref{alternativefits:FIG} for the O(4)-model and $L=16$. Both fits govern the data well in their fit ranges and provide information on the peak height of one particular volume. The right plot of fig.\,\ref{alternativefits:FIG} shows the extraction of $\nu$ with respect to \cref{extractionofnu} with $\xi=0$. For comparison the result of the global fit is also shown. The values for $\nu$ extracted from the three methods do not agree perfectly but the systematics from different fit approaches can be addressed.
\vspace{-0.25cm}
\vspace{-0.8cm}

%
%
\section{Results of the phase structure analysis}

Three different kinds of systematic effects have been addressed in the O(4)-model analysis of the critical exponent $\nu$. First, the previously mentioned variation of fit methods. Second, it was investigated how the absence of the largest and smallest volumes would effect $\nu$. Third, the fit ranges of the susceptibility fits were changed which turned out to be a small effect. The other two effects are of the order of the statistical error. In case of the HY model not enough data points with good enough statistics have been produced yet to perform a comparable analysis here.

\begin{figure}[H]
\begin{center}
\includegraphics[width=5.5cm]{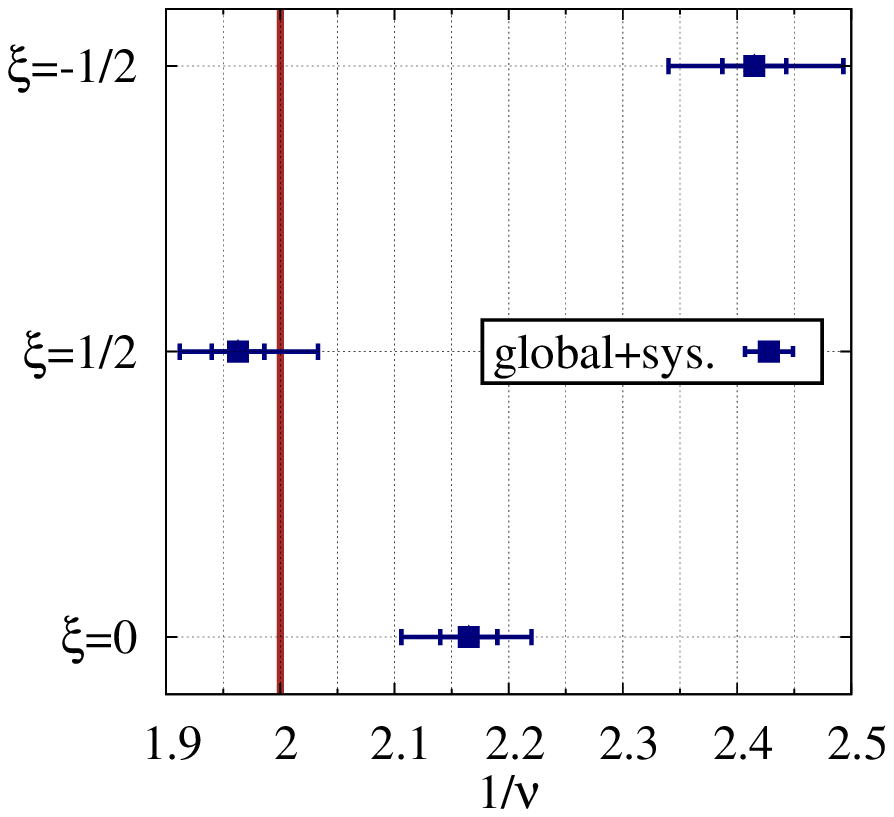}
\includegraphics[width=5.5cm]{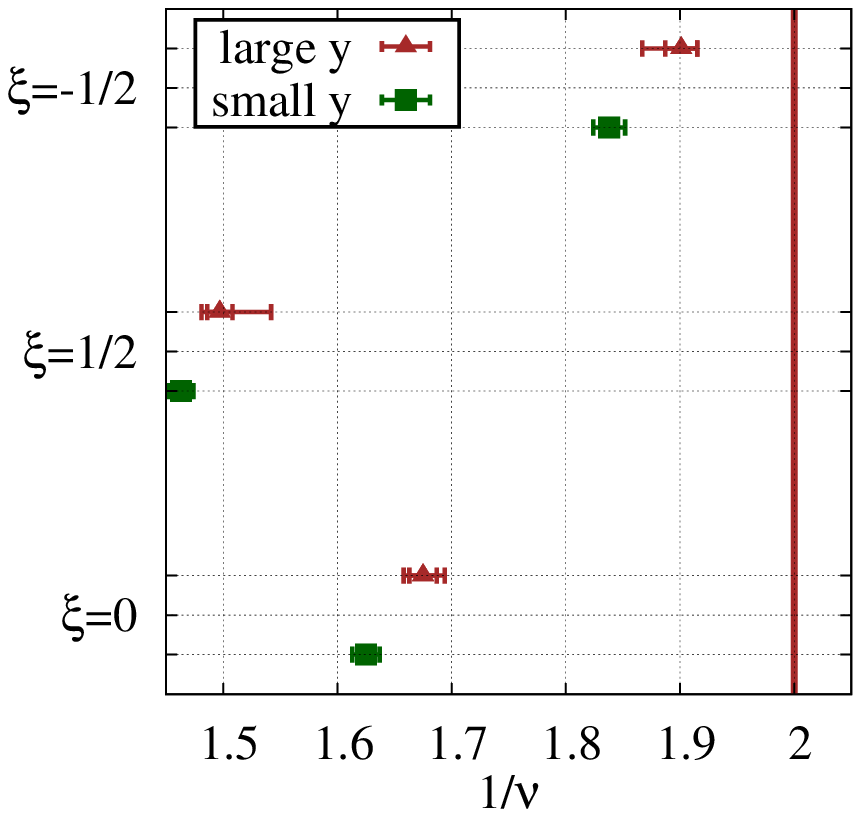}
\vspace{-0.4cm}
\caption{Results for the critical exponent $\nu$ in the O(4)-model {\it (left)} and the HY model {\it (right)} for various choices of logarithmic-correction exponents $\xi$.}
\label{finalnuanalysis:FIG}
\end{center}
\end{figure}

A comparison of final results for the O(4)-model and preliminary results for the HY model are shown in fig.\,\ref{finalnuanalysis:FIG}. Plotted is $1/\nu$ in dependence of the logarithmic-correction exponent $\xi$. The mean value is taken from the global fit with all volumes taken into account. The inner errorbars are statistical and the outer ones are statistical and systematic errors added in quadrature.

Three important observation can be made from fig.\,\ref{finalnuanalysis:FIG}. First, the inclusion of logarithmic corrections is important and has a big influence on the extraction of $\nu$. This becomes evident in the O(4)-model. Second, the sign of $\xi$ would be different in the HY model compared to the O(4)-model if the HY model was also trivial. Third, the $\nu$-dependence of $\xi$ is the same within errors for the phase transitions in the small and large Yukawa-coupling region. Hence, it stand to reason that both phase transition are in the same universality class. 
\vspace{-1.2cm}

%
%
\section{Spectrum calculations}

Not only the investigation of the phase transition of the HY model is interesting but also its spectrum. The easiest quantities to compute are the Higgs- and Goldstone boson masses, the field-renormalisation constants, and the fermion masses. The boson masses and the corresponding field-renormalisation constants can be computed from the momentum dependence of propagators \cite{Gerhold:2010wv}.
The masses presented in the following are based on a fit to the propagators according to a perturbative one-loop motivated expression \cite{Gerhold:2010wy} which allows a simultaneous extraction of masses and field-renormalisation constants. 
 
\begin{figure}[H]
\begin{center}
\includegraphics[width=5.3cm]{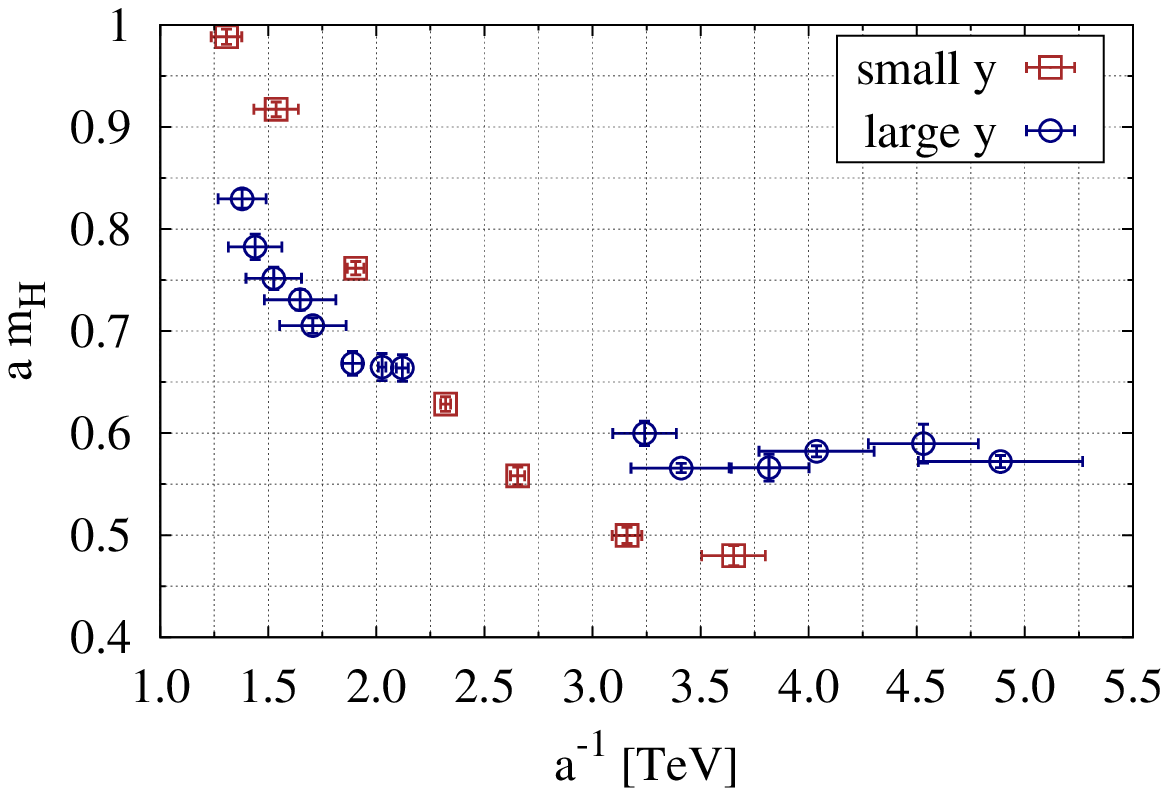}
\hspace*{-0.6cm}
\includegraphics[width=5.3cm]{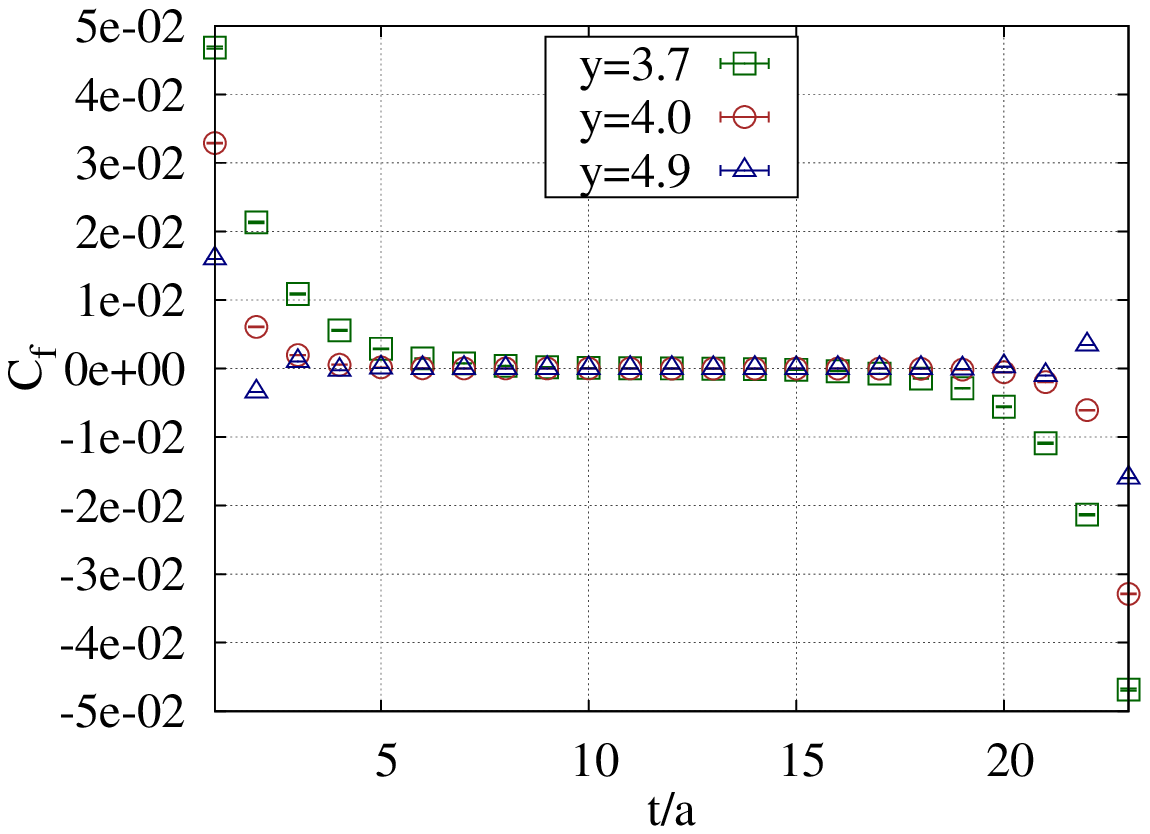}
\hspace*{-0.6cm}
\includegraphics[width=5.3cm]{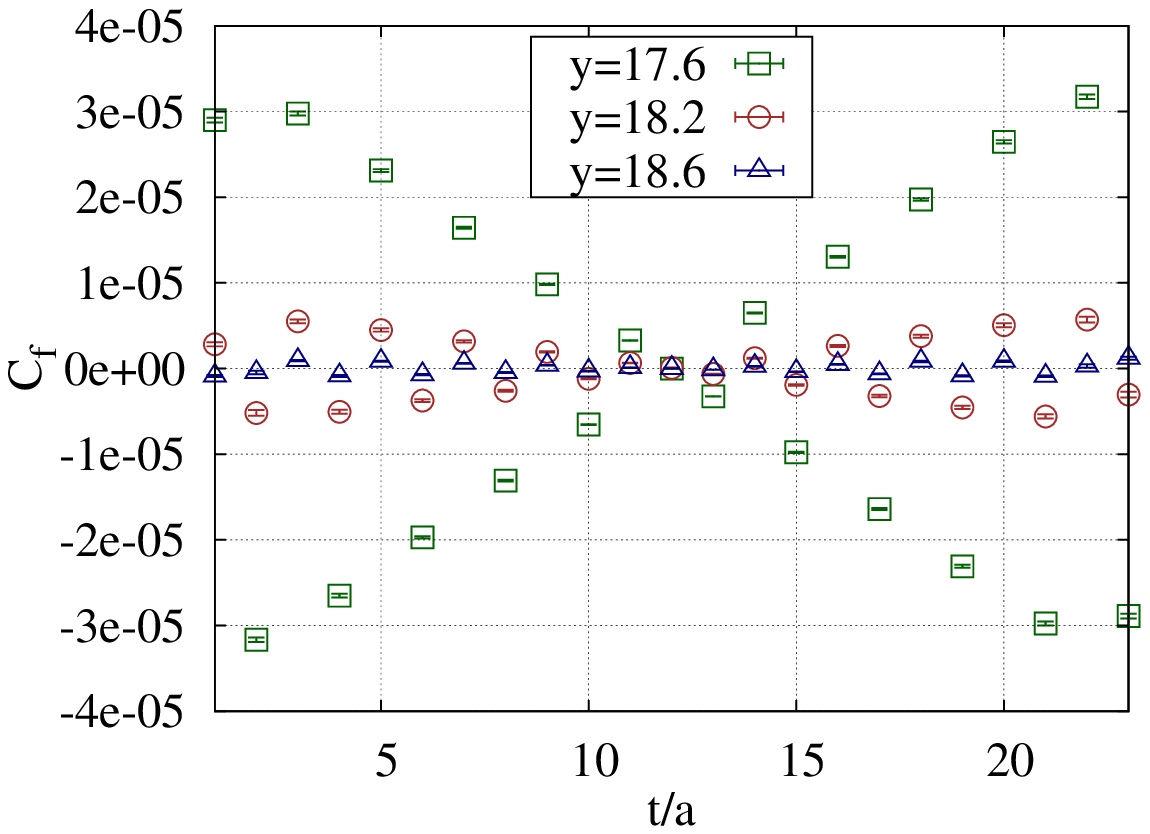}
\vspace{-1.0cm}
\caption{Higgs boson masses for $L=12$ in dependence of the cutoff for small and large Yukawa couplings {\it (left)}. Fermion correlation functions in the small {\it (middle)} and large {\it (right)} bare Yukawa coupling regime.}
\label{higgsmass:FIG}
\end{center}
\end{figure}

The spectrum investigation is still at an early stage and the Higgs boson masses are computed only for one volume $(L=12)$ so far. In the left plot of fig.\,\ref{higgsmass:FIG} the Higgs boson mass in lattice units is shown in dependence of the inverse lattice spacing
\begin{equation}
a = \frac{v_r}{246\text{ GeV}}, \quad v_r = \frac{v}{\sqrt{Z_G}},\quad v = \sqrt{2\kappa} \left< m_L \right> .
\end{equation}
for small and large Yukawa couplings. Although no volumes effects have been taken into account the Higgs boson mass dependence on the inverse lattice spacing is comparable in both regimes.
\vspace*{-0.2cm}
\vspace*{-0.6cm}

Examples of fermion correlation functions, from which the masses can be derived, are show in the middle and right plot of fig.\,\ref{higgsmass:FIG} for small and large Yukawa couplings. The fermion correlation functions are different close to the two phase transitions. In the small Yukawa coupling region (middle plot) the correlation function shows the expected $\sinh$-behaviour. In the large Yukawa coupling region (right plot) the correlation functions jumps from negative to positive values. This is a strong indication for doublers. Doublers are additional poles in the propagator which can be avoided by shifting their masses close to the cutoff \cite{WIlson1975}. In the large Yukawa-coupling region the fermion might become so heavy that it also gets a mass close to the cutoff and the doublers are not separated anymore. This effect does not vanish at a larger volume $(L=24)$ and higher cutoff $(\approx 8)$ GeV. Hence, this might a generic effect of the model and not a simple volume effect.
\vspace{-0.4cm}

%
%
\section{Conclusions and outlook}
\vspace{-0.2cm}
The analysis of the phase structure of this HY model is close to be finished. However, the findings of our preliminary spectrum calculations seem not to be consistent with the phase structure study. The fermion correlation function shows a different behaviour in the regimes of small and large Yukawa couplings while both phase transitions seem to be in the same universality class. In this study we only investigated the anomalous dimension connected to the Higgs boson mass, $\nu$, but did not investigate the anomalous dimension connected to the Yukawa coupling, $\delta$. It is still possible that $\delta$ turns out to be different for both bulk phase transitions and hence, that the phase transitions are not in the same universality class. This will be investigated in a later study.

We will finish our analysis of the critical exponent $\nu$ by increasing statistics and adding more points. This will allow us to perform a thorough study of systematic effect like for the O(4)-model. The spectrum will be investigated on larger volumes to address volume effects which are expected to be large in case of the boson masses. We will also scrutinize the possible appearance of bound states of fermions and Higgs bosons.
\vspace{-0.4cm}

%
%
\section{Acknowledgement}
\vspace{-0.2cm}
Simulations have been performed at the SGI system HLRN-II at the HLRN supercomputing service Berlin-Hannover, the PAX cluster at DESY-Zeuthen, HPC facilities at National Chiao-Tung University and National Taiwan University, and the cluster system $\varphi$ at KMI in Nagoya University. This work is supported by Taiwanese NSC via grants 100-2745-M-002-002-ASP (Academic Summit Grant), 99-2112-M-009-004-MY3, 101-2811-M-033-008, and 101-2911-I-002-509, and by the DFG through the DFG-project Mu932/4-4, and the JSPS Grant-in-Aid for Scientific Research (S) number 22224003.
\vspace{-0.5cm}

%
%

\end{document}